\documentclass[prd,showpacs]{revtex4}

\usepackage{textcomp}
\usepackage{mathtools, ulem, graphicx}
\usepackage[export]{adjustbox}
\usepackage{sidecap}
\usepackage{verbatim}
\newcommand{\e}{\begin{equation*}\begin{aligned}}
\newcommand{\ee}{\end{aligned}\end{equation*}}

\newcommand{\en}{\begin{equation}\begin{aligned}}
\newcommand{\een}{\end{aligned} \end{equation}}

\newcommand{\p}{\partial}

\newcommand{\f}[2]{\frac{#1}{#2}}

\newcommand{\ra}{\rangle}
\newcommand{\la}{\langle}

\newcommand{\da}{\dagger}
\newcommand{\ma}{\mathcal}

\newcommand{\tr}{\text{tr }}

\newcommand{\Q}{\left}
\newcommand{\W}{\right}

\newcommand{\pma}{\begin{pmatrix}}
\newcommand{\epma}{\end{pmatrix}}

\newcommand{\na}{\nabla}
\newcommand{\de}{\delta}

\newcommand{\Tr}{\text{tr}}

\begin{document}

\title{Bose and Fermi Statistics and the Regularization of the Nonrelativistic Jacobian for the Scale Anomaly}
\author{Chris L. Lin}
\author{Carlos R. Ord\'{o}\~{n}ez}
\affiliation{Department of Physics, University of Houston, Houston, TX 77204-5005}

\date{\today}
\email{cllin@uh.edu}
\email{cordonez@central.uh.edu}

\begin{abstract}
We regulate in Euclidean space the Jacobian under scale transformation for two-dimensional nonrelativistic fermions and bosons interacting via contact interaction and compare the resulting scaling anomalies. For fermions, Grassmannian integration inverts the Jacobian: however, this effect is cancelled by the regularization procedure and a result similar to that of bosons is attained. We show the independence of the result with respect to the regulating function, and show the robustness of our methods by  comparing the procedure with an effective potential method using both cutoff and $\zeta$-function regularization. 
\end{abstract}


\maketitle

%For the scale transformation 
%
%\begin{flalign}
%&\vec{x}\,'=\lambda \vec{x}, \\ \nonumber
%&t'=\lambda^2 t,\\ \nonumber
%&\psi'(\vec{x}\,' ,t')=\lambda^{-D/2} \psi(\vec{x},t). \nonumber
%\end{flalign}

%Setting $\lambda=1+\eta$ for infinitesimal $\eta$:

\section{Introduction}

The possibility of measuring the effects of quantum anomalies in nonrelativistic systems has prompted the development of quantum-field-theoretical approaches in the mathematical description of such anomalies \cite{hoff,brat,15,22,21}. In particular, we have recently developed a path-integral, Fujikawa approach to the calculation of anomalous corrections to virial theorems and equations of state for systems with a classical $SO(2,1)$ symmetry \cite{1,2,3,4,5}. Central to this approach is the ability to calculate the Fujikawa Jacobian $J$ for two dimensional Bose and Fermi particles with a contact interaction:

%\cite{hoff,brat,15,22,21}

\en \label{i1}
\ma L=\psi^*\Q(i\p_t+\f{\nabla^2}{2}\W)\psi-\f{\lambda}{2} (\psi^* \psi)^2,
\een

\en \label{i2}
\ma L=\sum\limits_{\sigma=\uparrow \downarrow}\psi^*_{\sigma}\Q(i\p_t+\f{\nabla^2}{2}\W)\psi_{\sigma}-\lambda \psi^*_{\uparrow}\psi^*_{\downarrow}\psi_{\downarrow}\psi_{\uparrow}.
\een

In \eqref{i1} and \eqref{i2}, the fields $\psi$ obey Bose and Fermi statistics respectively.
In \cite{2}, the path integral for the bosonic system was calculated for both zero and finite temperature, and the anomaly so calculated coincided with those obtained by other means in the literature \cite{norway,bergman}. These anomalies for systems with contact interactions control the anomalous sector for 2D, trapped ultracold dilute atoms, and, as shown by the author of \cite{hoff}, they can be interpreted as the Tan contact term, which determines much of the thermodynamics of such systems.  The results and formalism of \cite{1,2,3,4,5} for bosonic fields still remained to be developed for Fermi fields. We do so below, and we find similar results. In section II we give a short review of the Fujikawa approach for systems with classical scale invariance (more generally $SO(2,1)$), such as the ones studied here. The details of the calculation of the Jacobian $J$ are given in sections III and IV for the Fermi and Bose cases respectively. To further elucidate the consistency and robustness of our calculations, in section \ref{sec5} we compare for the fermion case the methods and results of this paper with an effective potential method using cutoff and $\zeta$-function regularization. The selection of the regulating matrix $M$ in both cases (fermionic and bosonic) is highlighted and the similarities and differences are commented upon in the conclusions.

\section{Scale Invariance and Fujikawa}

Under dilation the coordinates $x=(x_0,\vec{x})$ and fields $\phi_i$ transform as:

\en \label{1}
\vec{x'}&=e^{\eta} \vec{x}\\
x_0'&=e^{2\eta} x_0\\
\phi_i'(x')&=e^{[\phi_i]\eta}\phi_i(x),
\een

where $[\phi_i]$ is the length dimension of $\phi_i$ (in units where $\hbar=m=1$) which in two spatial dimensions is $[\phi_i]=-1$. Taking $\eta$ infinitesimal

\en \label{2}
\de \vec{x} &=\eta \vec{x}, \\
\de x_0 &=2\eta x_0,\\
\de \phi_i&=\eta \theta \phi_i,  \\
\theta &\equiv \Q(-1-\vec{x}\cdot \vec{\na}-2x_0 \p_{x_0} \W).
\een

A scale-invariant Lagrangian transforms as $\ma L'(x_0',\vec{x'})=e^{[\ma L] \eta}\ma L(x_0,\vec{x})$ where in two dimensions $[\ma L]=-4$

\en \label{3}
\delta \ma L&= \eta \Q(-4-\vec{x}\cdot \vec{\na}-2x_0 \p_{x_0} \W) \ma L=\eta \p_\mu \Q(\ma L f^\mu \W),\\
f^\mu&=(-2x_0, -\vec{x}).
\een

Under a change of variables $\phi'_i(x)=\phi_i(x)+\eta(x) \delta\phi_i(x)$ \footnote{$\delta \phi_i(x)$ here does not include the parameter $\eta(x)$.} the path integral becomes:

\en \label{4}
\int \prod_i d\phi_i \,e^{iS[\phi_i]}&=\int \prod_i d\phi'_i \,J^{\pm 1} e^{iS[\phi'_i-\eta(x)\delta\phi'_i(x)]}\\
&=
\int \prod_i d\phi'_i \,e^{-\int d^3x\, \eta(x) \ma A(x)} e^{i S[\phi'_i]-i\int d^3x\, \ma \eta(x) \delta L-i\int d^3x  \f{\p \ma L}{\p \p_\mu \phi'_i}  \p_\mu \eta(x) \de \phi'_i }\\
&=\int \prod_i d\phi'_i \,e^{i S[\phi'_i]} e^{-\int d^3x \,\eta(x) \ma A(x)+i \int d^3x\, \eta(x)\p_\mu \Q(\f{\p \ma L}{\p \p_\mu \phi'_i} \de \phi'_i-\ma L f^\mu \W)},
\een

where $+$ is for bosons, $-$ for fermions. With the far right term identified as the Noether current $j^\mu$, and noting that the field $\phi'$ is a dummy variable, for this to be true for arbitrary $\eta(x)$ then:

\en \label{5}
\p_\mu \la  j^\mu\ra =-i\la \ma A \ra.
\een

The anomaly $\la \ma A \ra$ which is given by the Jacobian in Fujikawa's method is:

\en \label{6}
\la \ma A \ra =\Tr \Q[\f{\de \de \phi_i(x)}{\delta \phi_j(y)}\W]_{x=y}=\pm \Tr\Q[\delta_{ij}\theta \delta(x-y)\W]_{x=y}.
\een

\section{Fermions}\label{fermions}

For the infinitesimal scale transformation

\en \label{7}
\de \vec{x} &=\eta \vec{x}, \\
\de t &=2\eta t,\\
\de \psi_{\uparrow\downarrow} &=\eta \theta  \psi_{\uparrow\downarrow}(\vec{x},t),\\
\de \psi^\da_{\uparrow\downarrow} &=\eta \theta  \psi^\da_{\uparrow\downarrow}(\vec{x},t),\\
\theta &\equiv \Q(-1-\vec{x}\cdot \vec{\na}-2t \p_t \W),
\een

we can apply Noether's theorem to the BCS Lagrangian $\ma L=\sum\limits_{\sigma=\uparrow \downarrow}\psi^\dagger_{\sigma}(i\p_t+\f{\nabla^2}{2})\psi_{\sigma}-\lambda \psi^\da_\uparrow \psi^\da_\downarrow \psi_\downarrow \psi_\uparrow$ which in two dimensions is classically scale-invariant, to get a conserved charge \cite{hoff}:

\en \label{8}
D=\int d^2 \vec{x} \, \vec{x}\cdot \vec{j}-2tH,\\
\vec{j}=-\f{i}{2} \Q( \psi^\da \vec{\nabla} \psi-\vec{\nabla}\psi^\da  \psi \W).
\een

The classical scale-invariance, hence the conservation law, is spoiled by the presence of a quantum anomaly \cite{jack}. The fermionic anomaly in Euclidean space is given by $\ma A=-\Tr [\theta \delta^3(x)\delta_{ij} \big ]|_{x=0}$ which differs in sign to the bosonic anomaly due to the transformation properties of Grassmann integrals \cite{book1,book2,book3,book4}. The trace is over the internal space of the fields. \\

%Note that $\p_\mu \delta^3(x)$, being odd, vanishes as $x=0$, so one can make the substitution $\theta=-1$ in the evaluation of the trace. \\ %\cite{book2}, \cite{book3}, \cite{book4}. \\  

To regulate the ill-defined expression $\Tr [\theta \delta^3(x)\delta_{ij} \big ]|_{x=0}$, we first rewrite the Lagrangian using a constraining field $\phi$:

\en \label{9}
\ma L=\sum\limits_{\sigma=\uparrow \downarrow}\psi^*_{\sigma}\Q(i\p_t+\f{\nabla^2}{2}\W)\psi_{\sigma}+
\f{\phi^*\phi}{\lambda}+(\psi_\uparrow \psi_\downarrow)\phi+(\psi_\downarrow^* \psi_\uparrow^* )\phi^*,
\een

which has the classical solution $\phi=-\lambda \psi^*_\downarrow \psi^*_\uparrow$ \footnote{This is equivalent to using the Hubbard-Stratonovich transformation in the path integral.}.\\

The Lagrangian can be written compactly as

\en \label{10}
\ma L=\f{1}{2}\pma \psi^* _\uparrow &  \psi_\downarrow    \epma  M   \pma \psi_\uparrow \\  \psi^*_\downarrow \epma+\f{1}{\lambda}\phi^* \phi \\
M=\pma
-\p_\tau+\f{\nabla^2}{2} & -\phi^* \\
-\phi & -\p_\tau-\f{\nabla^2}{2}
\epma , 
\een

where a transformation to Euclidean space has been made, and anticommutivity of the fields was used. \\

%$\int [d\phi d\phi^*][d\psi d\psi^*]_{\uparrow \downarrow}e^{\int \ma L \,d^dx}=\int [d\psi d\psi^*]_{\uparrow \downarrow} e^{\int (\ma L_f-\lambda \psi^*_\uparrow \psi^*_\downarrow \psi_\downarrow \psi_\uparrow)d^dx }$.\\

In momentum space the quadratic operator $M$ takes for constant $\phi$ the form:

\en \label{11}
M=\pma
i\omega -\f{k^2}{2} & -\phi^* \\
-\phi & i\omega +\f{k^2}{2}
\epma
\een

so that $M^\da M$ takes the form:

\en \label{12}
M^\da M=\pma
\omega^2 + \xi^2(\vec{k})+\phi\phi^* & 0 \\
0 & \omega^2 + \xi^2(\vec{k})+\phi\phi^* 
\epma ,
\een

%\en
%M&=\pma
%i\omega + \sqrt{\xi^2(\vec{k})+\phi\phi^*} & 0 \\
%0 & i\omega - \sqrt{\xi^2(\vec{k})+\phi\phi^*} 
%\epma\\
%\xi(\vec{k})&=\f{k^2}{2}
%\een

where $\xi(\vec{k})=\f{\vec{k}^2}{2}$. We first write $-\Tr \Q[\theta \delta^3(x-y) I_4 \W]\Big|_{x=y}$, where $I_n$ is the $n \times n$ identity matrix, as $-2 \, \Tr \Q[\theta \delta^3(x-y) I_2 \W]\Big|_{x=y}$. We regulate this expression by instead calculating:

\en \label{13}
-2\,\Tr \Q[\theta f\Q(\f{M^\da M}{\Lambda^4}\W)\delta^3(x-y) I_2 \W]\Big|_{x=y},
\een

where $f$ has the property that $f(\infty)=0$ and $f(0)=I_2$, and we take the limit $\Lambda \rightarrow \infty$ at the end of the calculation. This will regulate large eigenvalues of $M^\da M$ when $\delta^3(x-y) I_2 $ is expanded via a completeness relation using the eigenbasis \footnote{While in this case $M^\da M$ is already diagonal, in general one can transform into the eigenbasis in which $M^\da M$ is diagonal.} of $M^\da M$ \cite{fuji1}.  Since 

\en \label{14}
f\Q(\pma \lambda_1 & 0\\ 0 & \lambda_2 \epma \W)=\pma f(\lambda_1) & 0\\ 0 & f(\lambda_2) \epma, 
\een

we have

\en \label{15}
-\Tr \Q[\theta f\Q(\f{M^\da M}{\Lambda^4}\W)\delta^3(x-y) I_2 \W]\Big|_{x=y}=-\int \f{d\omega}{2\pi}\f{d^2\vec{k}}{(2\pi)^2}\Tr \Q[\theta f\Q(\f{M^\da M}{\Lambda^4}\W) \W] \\
=2 \int \f{d\omega}{2\pi}\f{d^2\vec{k}}{(2\pi)^2}\Q(1+i\vec{x}\cdot\vec{k}-2ix_0\omega_0\W)f\Q(\f{ \omega^2 + \xi^2(\vec{k})+\phi\phi^*}{\Lambda^4} \W) \\
=2 \int \f{d\omega}{2\pi}\f{d^2\vec{k}}{(2\pi)^2}f\Q(\f{ \omega^2 + \xi^2(\vec{k})+\phi\phi^*}{\Lambda^4} \W)\\
=2\Lambda^4  \int \f{d\omega}{2\pi}\f{d^2\vec{k}}{(2\pi)^2}f\Q(\omega^2 + \xi^2(\vec{k})+\f{\phi\phi^*}{\Lambda^4} \W) .
\een

where $\delta^3(x-y)$ was expanded in a Fourier transform. We Taylor expand the integrand:

\en \label{16}
\Tr \Q[f\Q(\f{M^\da M}{\Lambda^4}\W)\delta^3(x) I_2 \W]\Big|_{x=0}
=2\Lambda^4  \int \f{d\omega}{2\pi}\f{d^2\vec{k}}{(2\pi)^2}f\Q(\omega^2 + \xi^2(\vec{k})\W)+2\Lambda^4  \int \f{d\omega}{2\pi}\f{d^2\vec{k}}{(2\pi)^2}f'\Q(\omega^2 + \xi^2(\vec{k})\W)\f{\phi \phi^*}{\Lambda^4}+O\Q[\f{(\phi\phi^*)^2}{\Lambda^4}\W] .
\een

%Making the substitution $d^2\vec{k}=\Omega_2 kdk=\Omega_2 d\xi$, where $\Omega_2=2\pi$, we get:

%\en
%\Tr \Q[f\Q(\f{M^\da M}{\Lambda^4}\W)\delta^3(x) I_2 \W]\Big|_{x=0}
%=2\Lambda^4 \Omega_2 \int \f{d\omega}{2\pi}\f{d\xi}{(2\pi)^2}f\Q(\omega^2 + \xi^2+\f{\phi\phi^*}{\Lambda^4} \W) 
%\een

%Due to the eveness of the integrand, we can extend the range of $\xi$ from $[0,\infty)$ to $(-\infty,\infty)$ by dividing by an overall factor of 2:

%\en
%\Tr \Q[f\Q(\f{M^\da M}{\Lambda^4}\W)\delta^3(x) I_2 \W]\Big|_{x=0}
%=\Lambda^4 \Omega_2 \int \f{d\omega}{2\pi}\f{d\xi}{(2\pi)^2}f\Q(\omega^2 + \xi^2+\f{\phi\phi^*}{\Lambda^4} \W) 
%\een

%Now we go into polar coordinates $r^2=\omega^2 + \xi^2$:

%\en
%\Tr \Q[f\Q(\f{M^\da M}{\Lambda^4}\W)\delta^3(x) I_2 \W]\Big|_{x=0}
%&=\Lambda^4 \Omega_2^2 \int \f{rdr }{(2\pi)^3}f\Q(r^2+\f{\phi\phi^*}{\Lambda^4} \W) \\
%&=\f{1}{2}\Lambda^4 \Omega_2^2 \int \f{dr^2 }{(2\pi)^3} \Q[f\Q(r^2\W)+f'(r^2)\f{\phi\phi^*}{\Lambda^4}+... \W]
%\een

%The first term in the Taylor expansion is independent of the interaction. The 2nd term is easily integrated:

%\en
%\Tr \Q[f\Q(\f{M^\da M}{\Lambda^4}\W)\delta^3(x) I_2 \W]\Big|_{x=0}=
%\f{1}{2 (2\pi)^3}\Lambda^4 \Omega_2^2 \f{\phi\phi^*}{\Lambda^4} f(r^2)|^\infty_0 = -\f{1}{2 (2\pi)^3}\Lambda^4 \Omega_2^2 \f{\phi\phi^*}{\Lambda^4} 
%\een

The first term is independent of the interaction. The second term is evaluated in \eqref{a1} of the appendix, while higher terms vanish in the $\Lambda \rightarrow \infty$ limit.\\

The anomaly is therefore:

\en \label{17}
\ma A=-\Tr [\delta^3(x-y)I_4]\big |_{x=y}=-2\Tr [\theta \delta^3(x)I_2]\big |_{x=y}\\
=4\Lambda^4 \Q(-\f{ \Omega_2^2}{4 (2\pi)^3} \f{\phi\phi^*}{\Lambda^4} \W)=-\f{\phi\phi^*}{2\pi}.
\een

Plugging in the classical solution for the auxillary field:

\en \label{18}
\ma A=-\f{(\lambda \psi^*_\downarrow \psi^*_\uparrow)(\lambda \psi^*_\downarrow \psi^*_\uparrow)^*}{2\pi}
=-\f{\lambda^2}{2\pi}\psi^*_\uparrow \psi^*_\downarrow \psi_\downarrow \psi_\uparrow.
\een

%Note that $\Tr [\delta^3(x)I_4]\big |_{x=0}=-\Tr [\theta \delta^3(x)I_4]\big |_{x=0}$, since our regulator $\M^\da M$ is even in  

\section{Bosons}

Similarly, for the boson case, a saddle point expansion of the action $\int d^3x\, \ma L  =\int d^3x\, \Q[\psi^*(i\p_t+\f{\nabla^2}{2})\psi-\f{\lambda}{2}(\psi^* \psi)^2\W] $ about the classical solution $\psi_{\text{cl}}$ gives for the bilinear piece:

\en \label{19}
\ma L=\f{1}{2}\pma \psi^* &  \psi    \epma  M   \pma \psi \\  \psi^* \epma, \\
M=\pma
-\p_\tau+\f{\nabla^2}{2}-2\lambda \psi^*_{\text{cl}} \psi_{\text{cl}} & -\lambda \psi_{\text{cl}} \psi_{\text{cl}} \\
-\lambda \psi^*_{\text{cl}} \psi^*_{\text{cl}} & \p_\tau+\f{\nabla^2}{2}-2\lambda \psi^*_{\text{cl}} \psi_{\text{cl}}
\epma.
\een

Following the same procedure as in section \ref{fermions}, the regulating matrix $M$ becomes in momentum space:

\en \label{20}
M=\pma
i\omega -\f{k^2}{2}-2 \lambda \psi^*_{\text{cl}} \psi_{\text{cl}} & -\lambda \psi_{\text{cl}} \psi_{\text{cl}}\\
-\lambda \psi^*_{\text{cl}} \psi^*_{\text{cl}} & -i\omega -\f{k^2}{2}-2 \lambda \psi^*_{\text{cl}} \psi_{\text{cl}}
\epma.
\een

The generic matrix 

\en \label{21}
\pma A & B \\ C & D \epma
\een

has eigenvalues 

\en \label{22}
\pma \f{1}{2}(A+D) + \sqrt{\f{(A-D)^2}{4}+BC} & 0 \\ 0 &  \f{1}{2}(A+D) - \sqrt{\f{(A-D)^2}{4}+BC} \epma.
\een

First multplying $M^\da$ and $M$ and then using \eqref{22} for the eigenvalues of $M^\da M$ \footnote{Since we are computing a trace, we can always use the eigenbasis of $M^\da M$ that produces Eq. \eqref{23}.} one gets

\en \label{23}
M^\da M &=\pma
\Q(\sqrt{\omega^2 + \xi^2(\vec{k})+A^2}+\lambda \psi^* \psi \W)^2 & 0 \\
0 & \Q(\sqrt{\omega^2 + \xi^2(\vec{k})+A^2}-\lambda \psi^* \psi \W)^2 
\epma, \\
A^2&=2\lambda k^2\psi^* \psi+4 \lambda^2(\psi^* \psi)^2.
\een

Therefore $\int \f{d\omega d^2k}{(2\pi)^3} \,\tr f\Q(\f{M^\da M}{\Lambda^4}\W) \equiv \int \f{d\omega d^2k}{(2\pi)^3} \,\tr g\Q(\sqrt{\f{M^\da M}{\Lambda^4}}\W)$

\en \label{24}
&=\Lambda^4 \int \f{d\omega d^2k}{(2\pi)^3} \, \Q(  g\Q(\sqrt{\omega^2 + \xi^2(\vec{k})+\f{\tilde{A}^2}{\Lambda^2}}+\f{\lambda \psi^* \psi}{\Lambda^2} \W)
+g\Q(\sqrt{\omega^2 + \xi^2(\vec{k})+\f{\tilde{A}^2}{\Lambda^2}}-\f{\lambda \psi^* \psi}{\Lambda^2} \W) 
\W)\\
&=\Lambda^4 \int \f{d\omega d^2k}{(2\pi)^3} \, \Q[ 2g\Q(\sqrt{\omega^2 + \xi^2(\vec{k})+\f{\tilde{A}^2}{\Lambda^2}}\W)
+g''\Q(\sqrt{\omega^2 + \xi^2(\vec{k})+\f{\tilde{A}^2}{\Lambda^2}}\W)\Q(\f{\lambda \psi^* \psi}{\Lambda^2}\W)^2
\W]+O\Q[\f{(\psi^* \psi)^4}{\Lambda^4}\W],\\
&\tilde{A}^2=2\lambda k^2\psi^* \psi+\f{4 \lambda^2(\psi^* \psi)^2}{\Lambda^2}.
\een

The first term in the integrand can be rewritten as $f\Q(\omega^2 + \xi^2(\vec{k})+\f{\tilde{A}^2}{\Lambda^2} \W)$ which can be Taylor expanded:

\en \label{25}
f\Q(\omega^2 + \xi^2(\vec{k})\W)+f'\Q(\omega^2 + \xi^2(\vec{k})\W)\f{\tilde{A}^2}{\Lambda^2}+\f{1}{2}f''\Q(\omega^2 + \xi^2(\vec{k})\W)\Q(\f{\tilde{A}^2}{\Lambda^2}\W)^2\\
=f\Q(\omega^2 + \xi^2(\vec{k})\W)+f'\Q(\omega^2 + \xi^2(\vec{k})\W) \f{2\lambda k^2\psi^* \psi}{\Lambda^2}+f'\Q(\omega^2 + \xi^2(\vec{k})\W)\f{4\lambda^2(\psi^* \psi)^2}{\Lambda^4}+\\ 
\f{1}{2}f''\Q(\omega^2 + \xi^2(\vec{k})\W)\f{(2\lambda k^2\psi^* \psi)^2}{\Lambda^4}+O\Q[\f{(\psi^* \psi)^3}{\Lambda^6} \W].
\een

The first term is independent of the interaction and can be ignored. The second term can be renormalized into a chemical potential which explicitly breaks scale-invariance \cite{coleman}. Using \eqref{a1} of the appendix, the third and fourth integrals add to zero. \\

The integral of the last term in \eqref{24} is from \eqref{a1}:

\en \label{26}
\Tr \Q[\theta \delta^3(x) I_2 \W]\Big|_{x=0}=-\int \f{d\omega d^2k}{(2\pi)^3} \,\tr f\Q(\f{M^\da M}{\Lambda^4}\W)\\
=-\Lambda^4 \int \f{d\omega d^2k}{(2\pi)^3} \, g''\Q(\sqrt{\omega^2 + \xi^2(\vec{k})+\f{\tilde{A}^2}{\Lambda^2}}\W)\Q(\f{\lambda \psi^* \psi}{\Lambda^2}\W)^2\\
=-\Lambda^4 \int \f{d\omega d^2k}{(2\pi)^3} \, g''\Q(\sqrt{\omega^2 + \xi^2(\vec{k})}\W)\Q(\f{\lambda \psi^* \psi}{\Lambda^2}\W)^2+O\Q[\f{(\psi^* \psi)^3}{\Lambda^2}\W]\\
=-\f{\lambda^2(\psi^* \psi)^2}{4 \pi}.
\een

\section{Relationship with the Effective Potential Method} \label{sec5}

Fujikawa's method identifies the anomaly as the ill-defined expression $-\Tr [\delta^3(x-y)I_4]\big |_{x=y}$, which is regulated by $-2\, \Tr \Q[f\Q(\f{M^\da M}{\Lambda^4}\W)\delta^3(x-y) I_2 \W]\Big|_{x=y}$. It should be emphasized that $f$ is arbitrary except for the reasonable boundary conditions $f(0)=1$, $f(\infty)=f'(\infty)=0$. We will now specialize to $f(X)=e^{-X}$ to make a connection between Fujikawa's method and the effective action in the fermion case. Indeed, 

\en \label{29}
\Tr \Q[e^{-\f{M^\da M}{\Lambda^4}}\delta^3(x-y) I_2\W] \Big|_{x=y}=\Q\la x \Q| \Tr \text{ } e^{-\f{M^\da M}{\Lambda^4}} \W|x \W \ra \equiv h(x,x) 
\een 

is the heat kernel of $M^\da M$, where $M$ is the Hessian of Eq. \eqref{10}, and it should be kept in mind that $h(x,x)$ depends on the ``proper time'' $\f{1}{\Lambda^4}$. In what follows we will calculate Eq. \eqref{29} and use $\zeta$-regularization to derive the effective potential and from this the anomaly via the $\beta$-function. With the help of an infrared regulator, we will then calculate Eq. \eqref{29} as a series similar to Eq. \eqref{16}, which is analogous to a Seeley-DeWitt expansion \cite{qg, seel}, and with a cutoff regulator show that the anomaly is coming from the $\phi^\da \phi$ sector of the effective potential as indicated by Fujikawa's method. Finally, we will comment on the calculation of the determinant of $M^\da M$ and taking the square root (which halves the one-loop effective action up to a phase) rather than $M$, which unlike $M^\da M$, is not positive-definite, a critical feature of regularization in Fujikawa's method. \\

Performing the path integral over the fermion fields in Eq. \eqref{10} gives: 

\en \label{30}
\int [d\psi_\sigma][d\psi^\da_\sigma][d\phi][d\phi^*]e^{-\int d^2xd\tau \, \Q[\f{1}{2}\pma \psi^* _\uparrow &  \psi_\downarrow    \epma  M   \pma \psi_\uparrow \\  \psi^*_\downarrow \epma+\f{1}{\lambda}\phi^* \phi \W]}=\int [d\phi][d\phi^*] e^{-\int d^2xd\tau \, V_\text{eff}(\phi,\phi^*)},
\een

where $ V_\text{eff}(\phi,\phi^*)=\f{1}{\lambda}\phi^* \phi-\f{1}{VT} \ln\Q(\sqrt{\det \, M^\da M}\W)$. We will evaluate the determinant via construction of the $\zeta$-function. For constant $\phi$, using Eq. \eqref{12} on Eq. \eqref{29} we get:

\en
h(x,x)&=2\int\f{d\omega d^2\vec{k}}{(2\pi)^3}e^{-\f{\omega^2+\xi^2(\vec{k})+\phi^*\phi}{\Lambda^4}}\\
&=2\int\f{d\omega d\xi}{(2\pi)^2}e^{-\f{\omega^2+\xi^2+\phi^*\phi}{\Lambda^4}}=\f{\Lambda^4}{4\pi}e^{-\f{\phi^*\phi}{\Lambda^4}}.
\een

We construct the $\zeta$-function for $M^\da M$ \footnote{$\zeta(s)=\sum\limits_{i}\f{1}{\lambda_i^s}$, where $\lambda_i$ are the eigenvalues of $M^\da M$.} via analytic continuation using the Mellin transform \cite{Hawk}:

\en
\zeta(s)&=\f{1}{\Gamma(s)}\int^\infty_0 dt\, t^{s-1} \int d^3x \, h(x,x)\\
&=\f{1}{\Gamma(s)}\int^\infty_0 dt\, t^{s-1} \f{\mu^4}{4\pi t}e^{-\f{\phi^*\phi }{\mu^4}t} VT\\
&=\f{\mu^4}{4\pi}\Q(\f{\mu^4}{\phi^* \phi}\W)^{s-1} \f{VT}{s-1},
\een

where $VT$ is the volume of spacetime and $\f{1}{\Lambda^4}=\f{1}{\mu^4}t$ was introduced to make the proper time $t$ dimensionless \cite{ramon}, and $\mu$ is a momentum scale. Therefore we have:

\en \label{32}
\sqrt{\text{Det }M^\da M}&=e^{-\f{\zeta'(0)}{2}}\\
V_\text{eff}&=V_0+\f{1}{VT}\f{\zeta'(0)}{2}\\
&=-\f{\phi^* \phi}{\lambda(\mu)}+\f{\phi^*\phi}{8\pi}\Q(\log\Q(\f{\phi^*\phi}{\mu^4}\W)-1\W),
\een

where the tree level term $V_0$ comes from the Hubbard-Stratonovich transformation of Eq. \eqref{10}. Differentiating \eqref{32} w.r.t. $\mu$, the independence of $V_\text{eff}$ on $\mu$ gives $\beta(\lambda)=\f{\lambda^2}{2\pi}$. The anomaly is therefore \cite{pe}

\en \label{33}
\ma A=\beta(\lambda) \f{\p \ma L}{\p \lambda}=-\f{\lambda^2}{2\pi}\psi^*_\uparrow \psi^*_\downarrow \psi_\downarrow \psi_\uparrow,
\een

agreeing with Eq. \eqref{18}. \\

For the next method, instead of calculating the entire heat kernel, we will make the expansion indicated in Eq. \eqref{16}, and we are interested in the 2nd term on the RHS which in Fujikawa's method produces the anomaly. This term is:

\en
h_2(x,x)=2\Lambda^4  \int \f{d\omega}{2\pi}\f{d^2\vec{k}}{(2\pi)^2} e^{-\Q(\omega^2 + \xi^2(\vec{k})\W)}\f{\phi \phi^*}{\Lambda^4}.
\een

Because we are not summing the entire series, we will need to introduce an infrared regulator \cite{coleman} by making the replacement $\xi(\vec{k}) \rightarrow \xi(\vec{k})-\mu$, where $\mu$ is negative. This creates a positive gap in the spectrum which will help us avoid infrared divergences, and we will take $\mu \rightarrow 0$ at the end of the calculation.

\en
h_2(x,x)&=2\Lambda^4  \int \f{d\omega}{2\pi}\f{d^2\vec{k}}{(2\pi)^2} e^{-\Q(\omega^2 + \xi^2(\vec{k})-2\f{\mu \xi(\vec{k})}{\Lambda^2}+\f{\mu^2}{\Lambda^4} \W)}\f{\phi \phi^*}{\Lambda^4}\\
&=\f{1}{4\pi}\Q(1+\text{erf}\Q[\f{\mu}{\Lambda^2}\W]\W)\phi^*\phi,
\een

where $\text{erf}(x)=\f{2}{\sqrt{\pi}}\int_0^x dt\, e^{-t^2} $ is the error function. Using the identities $\ln \Q(M^\da M\W)=-\int^\infty_0 \f{e^{-M^\da M\epsilon}}{\epsilon} d\epsilon$ and $\sqrt{\text{Det }M^\da M}=e^{\f{1}{2}\int \Tr \ln \Q(M^\da M \W)d^3x}$ \cite{userguide} one gets:

\en \label{36}
V_\text{eff}&=-\f{\phi^* \phi}{\lambda(\Lambda')}-\f{1}{2}\int^\infty_\f{1}{\Lambda'^4} \f{h(x,x)}{1/\Lambda^4}      d\Q(1/\Lambda^4  \W)\\
V^{(2)}_\text{eff}&=-\f{\phi^* \phi}{\lambda(\Lambda')}-\f{1}{2}\int^\infty_\f{1}{\Lambda'^4} \f{\f{1}{4\pi}\Q(1+\text{erf}\Q[\f{\mu}{\Lambda^2}\W]\W)\phi^*\phi}{1/\Lambda^4}      d\Q(1/\Lambda^4  \W),
\een

where $\Lambda'$ was introduced to regulate the UV-divergence, which along with the condition $\mu <0$ for the IR, makes the integral convergent. Differentiating both sides of Eq. \eqref{36} w.r.t. $\Lambda'$ and using the fundamental theorem of calculus one gets:

\en  \label{37}
0&=\f{\lambda'}{\lambda^2}-\f{1}{2}\f{4}{\Lambda'^5} \f{\f{1}{4\pi}\Q(1+\text{erf}\Q[\f{\mu}{\Lambda'^2}\W]\W)}{1/\Lambda'^4}   \\
\Lambda' \lambda'&=\beta(\lambda)=\f{\lambda^2}{2\pi},
\een

where $\text{erf}\Q[\f{\mu}{\Lambda'^2}\W] \rightarrow 0$ both as $\mu \rightarrow 0$ and $\Lambda' \rightarrow \infty$. Therefore one can see that the anomaly comes from $h_2(x,x)$ when $h(x,x)$ is Seeley-DeWitt expanded in Eq. \eqref{36}. The reason that $h_2(x,x)$ (which is related to the 2nd term on the RHS of Eq. \eqref{16} using Fujikawa's method) determines the anomaly can be seen from a comparison of Eq. \eqref{32} with the 2nd term on the RHS of Eq. \eqref{36}: while $V_0=-\f{\phi^*\phi}{\lambda}$ is classically conformally invariant, $\f{\phi^*\phi}{8\pi}\Q(\log\Q(\f{\phi^*\phi}{\mu^4}\W)-1\W)$ is not due to the nonlocal term $\phi^*\phi \log\Q(\phi^*\phi\W)$ \cite{5}, and the behavior of this term is related to $\phi^*\phi \log[\mu^4]$, which is provided by the $h_2(x,x)$ term by performing the integral in Eq. \eqref{36}. \\

We now comment on setting $\sqrt{\text{Det }M^\da M}=\text{Det }{M}$. $M$ is Hermitian by itself in real space, but in Euclidean space, it is not, so use of $M^\da M$ was required in Fujikawa's method to expand $\delta^3(x-y)I_2=\sum\limits_{n} \phi_n(x)\phi_n^\da(y)$ in an eigenbasis $\phi_n(x)$ of $M^\da M$ and to regulate the eigenvalues with $f\Q(\f{M^\da M}{\Lambda^4}\W)$. Since $\sqrt{\text{Det }M^\da M}=\text{Det }{M}e^{i \theta}$, we lose the phase $\theta$ in the calculation of the effective potential, where $\theta$ is some real functional of the fields. However, in Euclidean space, this phase contributes an imaginary part to the effective action

\en
e^{-S_{\text{eff}}}=e^{-S_0+\ln\Q(\sqrt{\text{Det }M^\da M}\W)-i\theta}.
\een

While the possibility exists of complex effective potentials \cite{weinagain}, the $\beta$-function, being real, will not be affected by the addition of a complex part in Eqs. \eqref{32} and \eqref{36}, so the argumentation leading to Eqs. \eqref{33} and \eqref{37} would still be valid. As a check on this, the one-loop contribution to the effective potential can be written as \cite{Weinberg,bbs}

\en
V^{(1)}_{\text{eff}}=-\int \f{d^2\vec{k}}{(2\pi)^2} \Q[E(\vec{k})-\xi(\vec{k}\,) \W],
\een

where $E(\vec{k})=\sqrt{\xi^2(\vec{k})\,+\phi^*\phi}$ is the single-fermion excitation energy. Performing the integral with cutoff $\Lambda$ on the momentum gives 

\en
V^{(1)}_{\text{eff}}=\f{\phi^*\phi}{8\pi}\Q(\log\Q(\f{\phi^*\phi}{\Lambda^4}\W)-1\W)-\f{1}{4\pi}\Q(\phi^*\phi \W)\Q(\f{1}{2}+\ln 2\W),
\een

which agrees with the result of Eq. \eqref{32} after renormalization.

\section{Conclusions}

The fermion and boson anomalies for nonrelativistic scale-invariant systems such as those studied here have formally similar expressions $\pm (2)\Tr \Q[\theta \delta^3(x-y) I_2 \W]\Big|_{x=y}$, differing by a sign due to Berezin integration and a factor of 2 from the two fermion species, as expected. However, the trace is regulated with a different regulating matrix $M$ depending on the statistics. In both cases, the real time version of $M$ is Hermitian, but the Euclidean one is not, and hence we had to work with $M^\da M$ in order to assure the regulating effects of large eigenvalues of $M^\da M$ when $\delta^3(x-y) I_2 $ is expanded via a completeness relation. Our method reproduces known results for both fermions and bosons, Eqs. \eqref{18} and \eqref{26}. The robustness of our approach was studied by comparing our methods and results with an effective potential calculation using both cutoff and $\zeta$-function regularization.\\

%The result is that the anomalies for both systems are the same sign, $2\ma E-2P=-\f{\lambda^2}{2\pi}\psi^*_\uparrow \psi^*_\downarrow \psi_\downarrow \psi_\uparrow$ for fermions \cite{hoff} and $2\ma E-2P=-\f{\lambda^2}{4\pi}(\psi^*\psi)^2$ for bosons \cite{norway,bergman}.

In this work, we only considered the homogeneous case, i.e., constant background fields.  A heat kernel approach to consider non-homogeneous systems (trapped systems, for instance) is currently being developed, and we hope to report on this and applications to ultracold atoms elsewhere.

\begin{acknowledgements}

We thank the reviewer whose comments spurred the development of section \ref{sec5}. This work was supported in part by the US Army Research Office Grant No. W911NF-15-1-0445.

\end{acknowledgements}

\appendix*

\section{}

In this paper we make use of the following integrals:

\en \label{a1}
\int f'\Q(\omega^2+\xi^2(\vec{k})\W) d^2\vec{k} d\omega=-\f{\Omega_2^2}{4} \\
\int f''\Q(\omega^2+\xi^2(\vec{k})\W) k^4 d^2\vec{k} d\omega=\f{\Omega_2^2}{2}\\
\int f''\Q(\sqrt{\omega^2+\xi^2(\vec{k})}\W) d^2\vec{k} d\omega=\f{\Omega_2^2}{2}
 \een

where $f(0)=1$, $f(\infty)=f'(\infty)=0$, and $\Omega_2=2\pi$ is the two-dimensional solid angle.\\

A derivation is as follows:

\en \label{a2}
\int f^{(m)}\Q(\omega^2+\xi^2(\vec{k})\W)k^{4s} d^2\vec{k} d\omega=4^s \int f^{(m)}\Q(\omega^2+\xi^2(\vec{k})\W)\xi^{2s}  d^2\vec{k} d\omega
=4^s \int f^{(m)}\Q(\omega^2+\xi^2(\vec{k})\W)\xi^{2s}  d\omega \Omega_2kdk\\
=4^s \Omega_2 \int f^{(m)}\Q(\omega^2+\xi^2\W)\xi^{2s} d\omega d\xi 
\een

Due to the eveness of the integrand, we extend the integral over $\xi$ from $[0, \infty)$ to $(-\infty,\infty)$ by including a factor of 1/2, and then go into polar coordinates:

\en \label{a3}
\f{4^s \Omega_2}{2} \int f^{(m)}\Q(\omega^2+\xi^2\W)\xi^{2s} d\omega d\xi \\
=\f{4^s \Omega_2}{2} \int f^{(m)}\Q(r^2\W) r^{2s+1} dr \int \sin^{2s} \theta d\theta \\
=4^{s-1} \Omega_2\int f^{(m)}\Q(x \W) x^s dx \int \sin^{2s} \theta d\theta \\
\een

Plugging in $m=2$ and $s=1$, and $m=1$ and $s=0$, and integrating over $x$ by parts with the specified boundary conditions on $f$, gives the above two integrals. The third integral of \eqref{a1} proceeds similarly.

% with this new value of $M$, one finds that $\Tr [\delta^3(x)I_2]\big |_{x=0}=\f{\lambda^2}{4\pi} (\psi^\da \psi)^2$ [cf equation 30 of PRD paper]. However, although $\Tr [\delta^3(x)I_2]\big |_{x=0}$ differ in sign for bosons and fermions, the anomalies have the same sign, due to the fact that under a change of variables, bosons and fermion fields have inverse Jacobians from one another.

%\bibliography{fermionjack}
%\bibliographystyle{plain}

\end{document}